\documentstyle[twocolumn,prb,aps,epsfig]{revtex}

\begin{document}
\title{ Electronic and magnetic structure of CsV$_2$O$_5$ 
         }

\author{Roser Valent\'\i$^1$ and T. Saha-Dasgupta$^2$ } 

\address{$^1$Fakult\"at 7, Theoretische Physik,
 University of the Saarland,
66041 Saarbr\"ucken, Germany.}

\address{$^2$S.N. Bose National Centre for Basic Sciences, 
JD Block, Sector 3,
 Salt Lake City, Kolkata 700098, India.}

\date{\today}
\maketitle

\begin{abstract}
 We have studied the electronic structure of the spin-gapped
system CsV$_2$O$_5$ by means of an {\it ab initio} calculation.
 Our analysis and a re-examination of the susceptibility data
 indicate that the behavior of this system is  much closer
 to that
 of an alternating spin-$\frac{1}{2}$ antiferromagnetic chain with
 significant inter-dimer coupling and weaker inter-chain couplings
 than that of isolated
 dimers as was initially proposed. Comparison
 to the vanadate family members  $\alpha '$-NaV$_2$O$_5$, $\gamma$-LiV$_2$O$_5$
 and isostructural compounds like (VO)$_{2}$P$_{2}$O$_{7}$ is discussed.

\end{abstract}
PACS numbers: 75.30.Gw, 75.10.Jm, 78.30.-j 


\vspace*{1cm}

\section{Introduction}

 Low-dimensional spin-gapped quantum systems are of current interest
 since they show interesting ground states and a variety of unconventional
 low-lying
excitations.  Examples of such systems include spin-1 Haldane chains
 \cite{Haldane_83},
  spin-$\frac{1}{2}$ even-leg ladders\cite{Dagotto_96} and spin-$\frac{1}{2}$
 alternating chains \cite{Bonner_64}.  The discovery of appropriate
 compounds like Y$_2$BaNiO$_5$\cite{Darriet_93} (Haldane chain),
  Sr$_{14}$Cu$_{24}$O$_{41}$ \cite{Eccleston_98} or
 SrCu$_2$O$_3$ \cite{Azuma_94} (ladder systems) has brought new insight
into the study of these systems.
 There is a long list of various spin-$\frac{1}{2}$ alternating chain
systems
 which are being intensively studied in connection to their magnetic
 excitations, to mention a few of them KCuCl$_3$ \cite{Kato_98}, TlCuCl$_3$
\cite{Oosawa_01},
 Cu(NO$_3$)$_2$.$\frac{5}{2}$D$_2$O \cite{Xu_00}. 
 If frustration is also present 
new features appear in the magnetization spectrum of the
 low-dimensional quantum systems as in CuGeO$_3$ \cite{Hase_93},
(VO)$_2$P$_2$O$_7$ (VOPO) \cite{Garrett_97},
SrCu$_2$(BO$_3$)$_2$ \cite{Kageyama_99} 
 or in the recently synthesized Cu$_2$Te$_2$O$_5$Br$_2$ \cite{Lemmens_01}.

  An important family of low-dimensional compounds are the layered
 vanadates AV$_2$O$_5$ (A=Ca, Mg, Na, Li, Cs)\cite{Ueda_98}. While 
 CaV$_2$O$_5$ and
 MgV$_2$O$_5$ contain only magnetic V$^{4+}$ ions and 
 behave like spin-$\frac{1}{2}$ ladders with spin-gaps of the order
of 600K \cite{Onoda_96} and 20K \cite{Onoda_98}
respectively,  $\alpha '$-NaV$_2$O$_5$, $\gamma$-LiV$_2$O$_5$ and 
CsV$_2$O$_5$ are mixed-valence systems (V$^{4.5+}$ on
 average) with important charge and
spin fluctuations. This last property makes this family specially
 interesting since it allows to study the influence
 of charge-ordering and the corresponding crystallographic
 distortions on the magnetic interactions.
  The quarter-filled ladder compound $\alpha '$-NaV$_2$O$_5$
 \cite{Smolinski98,Horsch98}
 is a highly discussed
 material and
 has become a model substance for the study of spin, charge
and orbital coupling.  It behaves like a spin-$\frac{1}{2}$
 Heisenberg system at
 high temperatures and at 34K undergoes a charge ordering
 (2V$^{4.5+}$
$\rightarrow$ V$^{4+}$ + V$^{5+}$)
transition\cite{Ohama99} accompanied with a lattice distortion
  \cite{crystal} and the 
opening of a spin gap\cite{Isobe96}. $\gamma$-LiV$_2$O$_5$ is a charge ordered 
system with no spin gap opening at
low temperatures \cite{Isobe_96_2}.
  Recent investigations based on an {\it ab initio}
analysis of the electronic structure\cite{Valenti_01}
 show that charge fluctuations in this system are still significant
and the experimental
 observations \cite{Isobe_96_2,Takeo_99} can be well described by
 a  spin-$\frac{1}{2}$ asymmetric
quarter-filled ladder model. 
 Such proposal has also found support by a numerical analysis of
optical conductivity experiments\cite{Hubsch_01}.
 The third member of this vanadate family, CsV$_2$O$_5$, has been given
much less attention. Existing  susceptibility measurements of 
Isobe {\it et al.}\cite{Isobe_96_2}
could be rather well explained by considering this material as
a system of isolated dimers with a spin-gap of $\approx 12meV$. 
  Measurements other than
 that are unknown to us except for older susceptibility
measurements \cite{Mur_85} which somewhat disagree with  
  Isobe {\it et al.}'s
  data
 and different values of the exchange integral $J$
 and the gyromagnetic ratio $g$ are predicted.

 There are various reasons why we believe that
this system  deserves attention:
 (i) the question of how the electronic and magnetic
properties of this material compare
with the properties of the other two members of the alkali-vanadate 
family, {\it i.e.} $\alpha '$-NaV$_{2}$O$_{5}$ and $\gamma$-LiV$_{2}$O$_{5}$ 
  (ii) this system crystallizes  in a similar monoclinic 
      structure as the most discussed VOPO
and many  other materials  {\it e.g.} KCuCl$_3$,
TlCuCl$_3$ showing an alternating chain 
behavior of the magnetic interactions. It is then
 worthwhile to find out whether CsV$_2$O$_5$ has also a similar
behavior of the magnetic interactions as its isostructural members.
 (iii)  as will be shown below, 
a re-examination of the susceptibility measurements reveals that
 a dimer model is not the only model that  can provide
reasonable fitting of the susceptibility data.

  In view of the above, we have performed 
 a microscopic study of this system  by  considering
 Density-Functional-Theory-based  (DFT) calculations.
  The microscopic study of the electronic behavior of
 this compound reveals that the inter-dimer coupling
 in this material in addition to the intra-dimer interaction is 
 also significant and shouldn't be neglected.
  Further, an analysis of the susceptibility data shows that
 getting estimates of the coupling constants and the gyromagnetic factor
based merely on a fitting procedure is very subtle and with such an 
a\-na\-lysis one can only test the consistency with the assumed model,
 but cannot prove the uniqueness of that model. Results on that will
 be presented in the next sections. We shall also discuss
 the similarities and differences to the other members of
 this vanadate family, $\alpha '$-NaV$_{2}$O$_{5}$
 and $\gamma$-LiV$_{2}$O$_{5}$ as well
 as to isostructural compounds like VOPO.
 
\section {Crystal Structure}
\begin{figure}[t]

\vspace*{0.1cm}

\centerline{\hspace*{1.5cm}
\epsfig{file=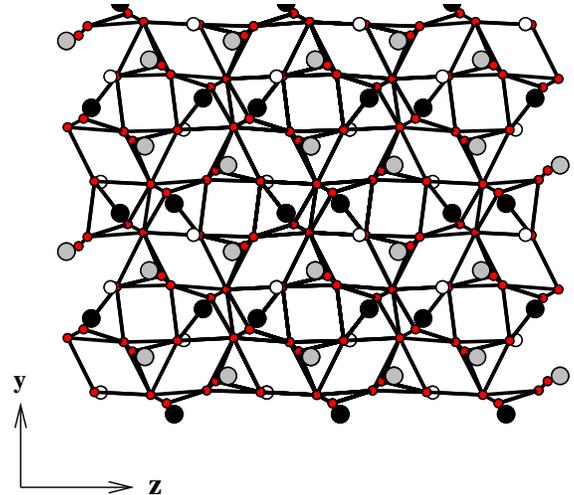,width=0.45\textwidth}
           }
\caption{\label{struct_yz}
Crystal structure of CsV$_2$O$_5$ projected in the $(yz)$ plane (see text
for explanation). 
  The large
circles are the V-ions, black and grey for V$^{4+}$ and V$^{5+}$ respectively.
The oxygens are represented by the smaller
circles. Pairs of edge-shared distorted V$^{4+}$O$_5$ square pyramids
 are bridged by V$^{5+}$O$_4$ tetrahedra to form  layers.
 The alkali-ion Cs 
shown by white circles , are located in between these sheets.}
\end{figure}

 The crystal structure of CsV$_2$O$_5$ is
  somewhat different from that of $\alpha '$-NaV$_2$O$_5$ and
 $\gamma$-LiV$_2$O$_5$.
  $\alpha '$-NaV$_2$O$_5$ and $\gamma$-LiV$_2$O$_5$ crystallize in
 the orthorhombic space group $Pmmn$ ($D_{2h}^{13}$) and $Pnma$ 
($D_{2h}^{16}$) respectively.  They consist
 of a double chain-structure
 of edge-sharing distorted square VO$_5$ pyramids
running along the orthorhombic $y$ axis, which are linked together
 via common corners to form layers. These layers are stacked upon
 each other along $z$. Na/Li atoms are located between these layers.   
 CsV$_2$O$_5$ crystallizes\cite{Waltersson_77}, on
 the other hand, in the  monoclinic space group
  $P2_1/c$ ($C_{2h}^5$).  
CsV$_{2}$O$_{5}$ has also a layered structure with Cs ions in between the 
layers.
  These layers, which are somewhat tilted ( {\it i.e.} not
  strictly in {\it yz} plane), and are stacked
 upon each other along $x$,
 are made of two types of vanadium atoms V1 and V2. 
 The V1 atoms are in a distorted square
 pyramidal environment of oxygens and form pairs by sharing the edge
  of the pyramids.  These pairs are   
 separated from one another by the V2 vanadiums in 
  a tetrahedral co-ordination of oxygens 
 as shown in Fig.\ \ref{struct_yz}.  
   Within a pair
of square pyramids, the two apical O atoms are pointing in opposite
directions and  the remaining O atoms of the pyramids are shared with
  tetrahedra. The V1-V1 distance across the shared edge of the 
   square-pyramids-pair is 3.073 $\AA$ and the V1-V2 distance
   across the shared corner of a pyramid and a tetrahedron is 3.352 $\AA$.
  Since
 the average V2-O distance  within the tetrahedron d=1.718 $\AA$ is
 shorter than the average V1-O distance within the square pyramid d=1.882
  $\AA$, one can conclude from a bond-valence analysis \cite{Waltersson_77}
 that
  the V1 sites are  in the oxidation V$^{4+}$ and the V2
  sites are in the oxidation V$^{5+}$.   
  As the 
  intra-pair V$^{4+}$-V$^{4+}$ distance is 3.073 $\AA$ while
 the inter-pair distance along the $z$-axis is  5.501 $\AA$, this
 compound has been assumed to be a system of isolated dimers.
 
\section {Band Structure}
 In order to analyze the electronic behavior of CsV$_2$O$_5$,  we
 carried out DFT calculations within the Local Density
 Approximation (LDA) 
 by employing the full-potential
 linearized augmented plane wave code WIEN97 \cite{WIEN97} (LAPW) and by 
linear-muffin-tin orbital (LMTO)\cite{Andersen_75}
 based on the Stuttgart TBLMTO-47 code.  Both calculations agree well
 within the allowed
 error-bars of the various different approximations involved in
 two different methods. In both approaches we treated the 
exchange-correlation part by using the generalized gradient approximation
\cite{Perdew_96}.

\begin{figure}[t]
\epsfxsize=0.5\textwidth
\centerline{\epsffile{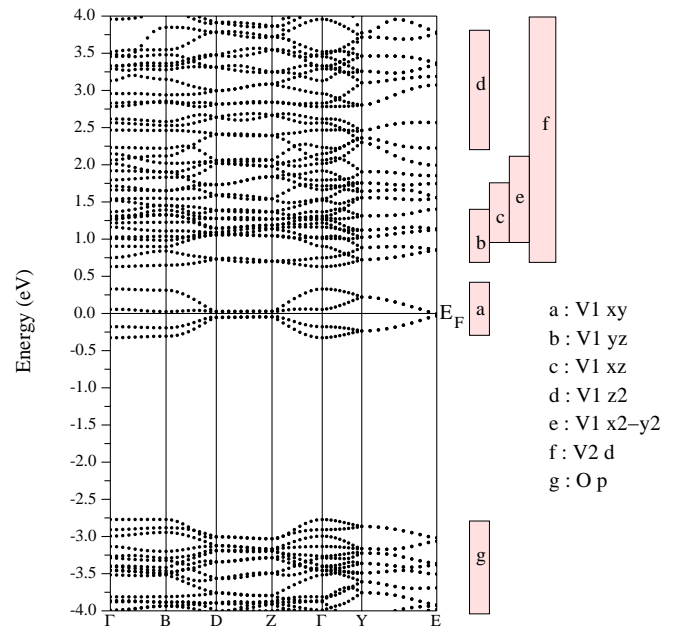}\hfill}

\vspace*{-0.1cm}
\caption{\label{Cs_spag}
DFT results for CsV$_2$O$_5$. The path is 
along $\Gamma$=(0,0,0), B=(-$\pi$,0,0),
D=(-$\pi$,0,$\pi$),
Z=(0,0,$\pi$), $\Gamma$,
Y=(0,$\pi$,0),
E=(-$\pi$,$\pi$,$\pi$). Also shown in rectangles is the band
character (in the
 local coordinate system) where V1=V$^{4+}$ and V2=V$^{5+}$.}
\end{figure}

\begin{figure}[t]

(a)
\epsfxsize=0.5\textwidth
\centerline{\epsfig{file=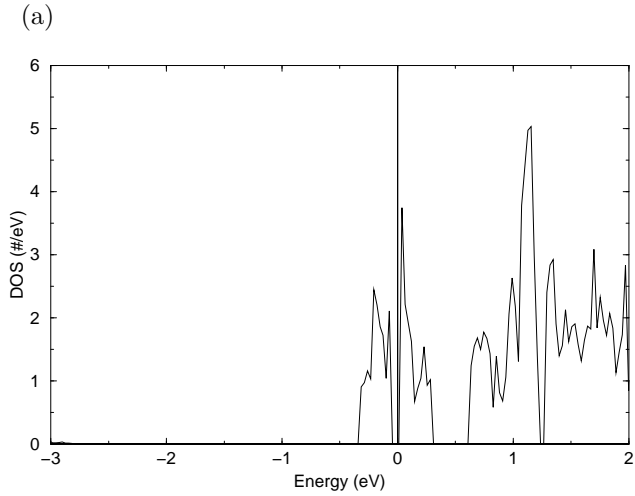 ,
width=0.33\textwidth,angle=-90}}

(b)

\epsfxsize=0.5\textwidth
\centerline{\epsfig{file=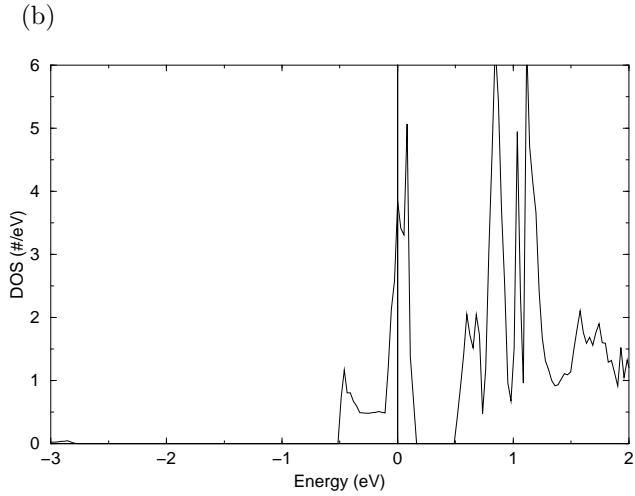 ,
width=0.33\textwidth,angle=-90}}

(c)

\epsfxsize=0.5\textwidth
\centerline{\epsfig{file=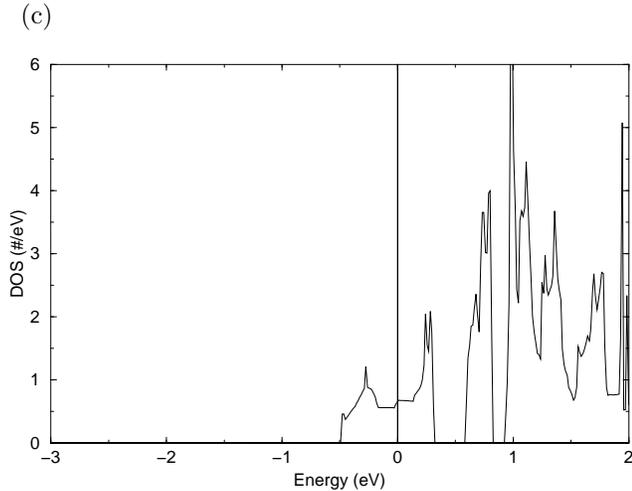 ,
width=0.33\textwidth,angle=-90}}

\vspace*{-0.1cm}
\caption{\label{partial_DOS}
Partial density of states of the V$^{4+}$-3d orbitals in (a) CsV$_2$O$_5$
 (b) $\gamma$-LiV$_2$O$_5$  (c) $\alpha '$-NaV$_2$O$_5$ 
obtained from DFT. Compare
 the distinct behavior at band edges around the Fermi level.
}
\end{figure}

In Fig.\ \ref{Cs_spag}
 the energy bands for CsV$_2$O$_5$ along the path $\Gamma 
BDZ\Gamma YE$
 are shown. Since the unit cell of CsV$_2$O$_5$ contains four
formula units, i.e. four V$^{4+}$ and four V$^{5+}$, 
there are four bands around the Fermi level which correspond
to V$^{4+}$-3d orbitals.  These bands are separated by an energy gap of
 about 2 eV from the lower O-p valence bands and an energy gap
of about 0.5 eV from the upper conduction bands. The conduction
 bands  up to 4eV are of V-3d nature.  There are a few
 points to  be noted here:
 (i) The energy range on the CsV$_2$O$_5$ bands
is of the same order of magnitude
    as it is in $\alpha '$-NaV$_2$O$_5$\cite{Smolinski98}
 and $\gamma$-LiV$_2$O$_5$\cite{Valenti_01}.
(ii) The contribution of V$^{5+}$-3d is mostly in
 the upper conduction bands, starting from about 0.75 eV above the Fermi level
 and
 have a very small contribution at the Fermi level indicating that,
contrary to $\alpha '$-NaV$_2$O$_5$ and $\gamma$-LiV$_2$O$_5$, 
here the system is close to complete
 charge ordering.
 (iii) There is no dispersion along the $x$ axis
 while along the $z$ axis (chain direction) the dispersion is 
non-negligible and of about 300meV. Along the $y$ axis the LDA-bands
 show a small dispersion which corresponds to inter-chain interactions.
In order to investigate the dimension-dependent behavior of
 CsV$_2$O$_5$, 
 we show  in Fig.\ \ref{partial_DOS}(a) the partial density of states (DOS)
of the V$^{4+}$ 3d-orbitals for CsV$_2$O$_5$.
 For comparison we have also plotted in Fig.\ \ref{partial_DOS}(b)
the partial density of states
of V1-3d orbitals for $\gamma$-LiV$_2$O$_5$\cite{Valenti_01} and
 in Fig.\ \ref{partial_DOS}(c) that of V-3d orbitals for
 $\alpha '$-NaV$_2$O$_5$. Note that, while the
 $\gamma$-LiV$_2$O$_5$ and $\alpha '$-NaV$_2$O$_5$ DOS
 show the characteristic quasi-1D van Hove
 singularities
near the band edges around the Fermi level,
 this is not the case for the Cs-compound. The DOS
for CsV$_2$O$_5$ shows more a 2D-behavior than 1D.
(iv) In the global frame of reference, the character of
 the bands around the Fermi level turn out to be that of d$_{z^{2}}$
 with small admixture from d$_{x^{2}-y^{2}}$ character. Rotation to
local co-ordinate system with the $z$-axis aligned along the 
{\it (V-apical O)} direction for individual pyramids transform this
to d$_{xy}$ character, in agreement with the crystal field analysis
of a V$^{4+}$ ion in the pyramidal environment\cite{Ueda_98}.

The band-structure analysis leads us to the conclusion that
 inter-dimer coupling is non-negligible. If
 the system  would have been a pure dimer system we would 
have expected no dispersion in any direction.
The existence of a non-negligible dispersion tells us that a
 model of unlinked dimers is too simple and there must be some
 inter-dimer coupling to explain the band picture.
 From the DOS
 analysis we learn that inter-dimer interactions in both $z$
and $y$ direction are important for the electronic properties.
The possible role of the V$^{5+}$O$_4$ tetrahedra in this context will be
 investigated in the next section.

 \section {Microscopic model (Effective model)} 
\begin{figure}[t]
\centerline{
\epsfig{file=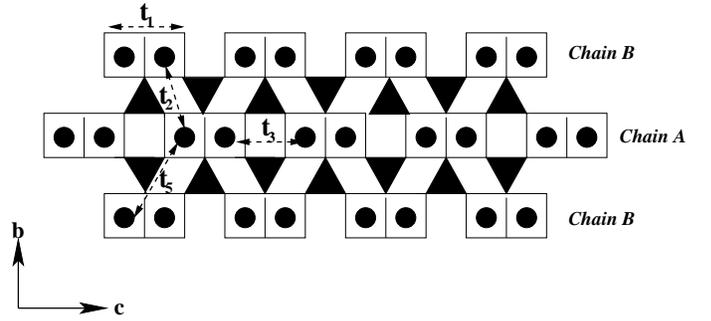,width=0.50\textwidth} 
           }
\vspace{4pt}
\caption{\label{hoppings}
Hopping parameters corresponding to the tight-binding modeling
of CsV$_2$O$_5$.  Not shown is the on-site energy 
$\varepsilon_0$ of the $V^{4+}$ site.  While $t_1$ corresponds
to a direct V$^{4+}$-O-V$^{4+}$ superexchange path, $t_2$, $t_3$ and
$t_5$ are defined by pathes through the neighboring V$^{5+}$O$_4$ tetrahedra.}
\end{figure}

\begin{figure}[t]
\centerline{
\epsfig{file=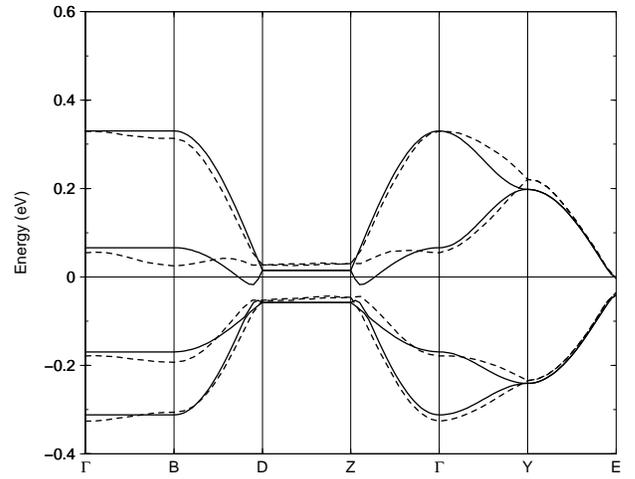,width=0.35\textwidth,angle=-90}
           }
\vspace{4pt}
\caption{\label{TB_fit}
Comparison of the tight-binding bands (solid lines)
with the DFT bands (dashed lines). The tight-binding parameters
 (see Fig.\ \protect\ref{hoppings}) are (in eV)
$\varepsilon_0=-0.0215$, $t_1=0.117$, $t_2=0.015$,
$t_3=0.097$, $t_5=0.050$.}
\end{figure}

In order to understand the nature of the ground state in 
CsV$_2$O$_5$  we need to determine the appropriate microscopic model
which explains the low-energy physics of this compound.
 We have therefore applied the so-called {\it downfolding
method}  offered by the new and generalized version of the
LMTO \cite{newlmto} method, together with a tight-binding (TB) analysis
on the band-structure results. 
The downfolding technique consists of  integrating out the
high energy degrees of freedom so as to describe the details of LDA
energy bands close to the Fermi energy in terms of few-orbital
effective Hamiltonians.
 From this analysis, we can extract the effective
hopping matrix elements $t_{ij}$ between vanadium ions in CsV$_2$O$_5$
 (by Fourier Transform of the downfolded Hamiltonian $H(k) \rightarrow
H(R)$ ) which reproduce the behavior of the  LDA bands close to the
Fermi level. Since, as seen in the previous section,
 the contribution of the V$^{5+}$-3d orbitals 
 around the Fermi level is very small, and the biggest contribution
comes from  the V$^{4+}$-3$d_{xy}$ orbital,
 a one-orbital effective tight-binding
Hamiltonian should give a good description of the low energy physics.
 The tight-binding Hamiltonian  therefore, can be written as:

\begin{eqnarray}
H_{TB}=-\sum_{<i,j>}t_{ij}(c^+_jc_i+c^+_ic_j)
\end{eqnarray}
where $i$ and $j$ denote a pair of V$^{4+}$ ions labeled each one
by $(n,s)$ where $n$ denotes the unit cell and $s=1,2,3,4$  V$^{4+}$
ions in the
unit cell.   In Fig.\ \ref{hoppings}  all the considered hoppings
 $t_1$, $t_2$, $t_3$ and t$_5$ have been drawn.  The subscript $1$, $2$,
 $3$ and $5$ indicate respectively that the hopping integral is between
 first nearest neighbors (n.n.), second n.n., third n.n and fifth n.n.
  The hopping term $t_4$, which connects two V$^{4+}$ ions
belonging to two different layers along $x$,
 has been neglected  due to its vanishingly
small value.    
  Note  here
 that no hopping terms have been considered along the $x$ direction,
 since as we have learnt from the band picture, there is no
 dispersion along this direction and that means that the 
 interlayer coupling along $x$ must be negligible.
 In Fig.\ \ref{TB_fit} we show a comparison of the
 {\it downfolding}-TB bands to the
 LDA bands.  We observe that not only the hopping corresponding 
 to the intra-dimer interaction $t_1$  is important but also
the inter-dimer matrix element $t_3$ 
 along $z$ is considerable. A model with only $t_1$ and $t_3$, i.e.
 alternating chain, shows a dispersion relation of the form:
\begin{eqnarray}
 E=\pm \sqrt{t_1^2 + t_3^2+2t_1t_3cosk_z}
\end{eqnarray} 
which is symmetric with respect to E=0 (i.e. to $E_F$) and doesn't
 account for the band-splittings in the paths  $\Gamma - D$ 
and $\Gamma - Y$. These
 features are described by the inter-chain
hopping parameters along the $y$ direction, $t_2$ and $t_5$,
though they are much smaller than $t_1$ and $t_3$. 
Note that  the TB-results show an artificial band-crossing along
 the path $B-D$ and $Z-\Gamma$.  We have analyzed this result and seen that
 the inclusion of longer-ranged interactions lifts this
artificial crossing.

Concerning the nature of the hopping integrals, we observe that 
 the values of the dominant hopping parameter  $t_1=0.12 eV$  
 (which agrees with the overall band-width of the V$^{4+}$-d$_{xy}$
 derived low-energy bands in CsV$_2$O$_5$) is smaller
 than the dominant hopping parameter
 for $\alpha '$-NaV$_2$O$_5$\cite{Smolinski98} $t_a=0.37 eV$
 and $\gamma$-LiV$_2$O$_5$\cite{Valenti_01} $t_a=0.35 eV$. This difference
 can be explained by the nature of the path. While $t_1$ in CsV$_2$O$_5$
 corresponds to a $V-O-V$ path in between two {\it edge}-sharing square
pyramids, $t_a$ in $\alpha '$-NaV$_2$O$_5$ and $\gamma$-LiV$_2$O$_5$
corresponds to a $V-O-V$ path between two {\it corner}-sharing
square pyramids. This feature has important implications in the
 strength of the exchange interaction as has been pointed out in
Ref.\ \cite{Ueda_98}.  Also note that the same kind of {\it edge}-sharing
 pyramids are also present in $\alpha '$-NaV$_2$O$_5$ and
$\gamma$-LiV$_2$O$_5$\cite{Valenti_01} with effective hopping values
smaller than $t_a$ and similar to $t_1$ in CsV$_2$O$_5$.

  About the nature  of the rest of
the hopping integrals in CsV$_2$O$_5$, we observe that $t_2$,
$t_3$ and $t_5$ correspond to V$^{4+}$-V$^{4+}$ paths through the bridging
 V$^{5+}$O$_4$ tetrahedra with different lengths and angles.
 Beltran {\it et al.}\cite{Beltran_89} 
studied, based on geometrical considerations, the
 various super-exchange paths for the case of vanadyl phosphates.  In
particular they considered the VOPO compound.  VOPO and CsV$_2$O$_5$
have a similar monoclinic structure if one identifies the P$^{5+}$O$_4$
tetrahedra and the V$^{4+}$O$_5$ square pyramids in VOPO with the V$^{5+}$O$_4$
tetrahedra and the V$^{4+}$O$_5$ square pyramids in CsV$_2$O$_5$
respectively.  As argued by Beltran {\it et al.},  based on the
formation of coherent molecular electron orbitals, the coupling
between two V$^{4+}$
through bridging oxygens for separated pyramids is weaker than
those via the pathways involving V$^{5+}$ ions. Furthermore,  double
bridging modes, involving two V$^{5+}$ ions, as in $t_2$ and $t_3$
 (see Fig.\ \ref{hoppings}) can be stronger than single bridging modes involving one V$^{5+}$
ion as in $t_5$.
 These interactions can be as strong as paths of the type
$t_1$, i.e. V$^{4+}$-O-V$^{4+}$ where the square pyramids are linked to each other
by sharing an edge. Geometry considerations then lead to a relative order of
  importance among the various bridging modes, which for
CsV$_2$O$_5$ is observed to be $t_{3} > t_{5} > t_{2}$. 
 We note here that while in CsV$_2$O$_5$ we have seen that the dimers
 are formed by the edge-sharing pyramids ($t_1$) with the
 predominant inter-dimer interaction formed by V$^{4+}$-V$^{4+}$ 
 interaction bridged by V$^{5+}$O$_{4}$ ($t_3$), in VOPO these roles
 are interchanged in the sense that dimers are formed by
 V$^{4+}$O$_5$ square pyramids bridged by P$^{5+}$O$_4$ groups\cite{Garrett_97,tanusri}
 ($t_3$) and the inter-dimer interaction is formed by
 neighboring V$^{4+}$O$_{5}$ pyramids ($t_1$). This is due to the different type
 of distortion in the square pyramids as well as different angle
 and distance  relations between these two compounds. 

 Using the relation $J_1=4 t_1^2/U$, $U$ being
 the on-site Coulomb electron-electron interaction and considering 
 U$\sim$ 2.8 eV \cite{Smolinski98}
 the DFT calculation gives an estimate of the intra-dimer exchange
 coupling, $J_1 \sim 225 K$.

To summarize, the conclusions that we draw from the above
 analysis is that {\it ab initio} calculations support
an alternating chain behavior for CsV$_2$O$_5$ along $z$ with weaker
inter-chain interactions along $y$. 
 
\section{Susceptibility} 

In order to test the
results obtained from the {\it ab initio} calculation we
 only have available two sets of susceptibility data \cite{Isobe_96_2,Mur_85}
 as a function
 of temperature  which
 don't completely agree quantitatively with each other.
 We consider here the data by Isobe {\it et al.}\cite{Isobe_96_2}
 These authors proposed that their experimental data are well fitted
 by the susceptibility of a spin-$\frac{1}{2}$ dimer system
with $H=J{\bf S}_1{\bf S}_2$ given by:
\begin{eqnarray}
\label{dimer_susc}
\chi_{raw}= \chi_{CsV_2O_5} + \chi_{imp} \\ \nonumber
  = \frac{Ng^2\mu_B^2}{k_BT}\frac{1}{3+\exp( J/k_BT)}
  + \chi_0 + \chi_{imp}
\end{eqnarray}
 with a g-factor of 1.8,  $ J=146K$ and
 $\chi_0=8\times 10^{-5}{emu}/{mol}$.
 We have re-analyzed the data by considering the temperature
dependence of the susceptibility
for a spin-$\frac{1}{2}$
 alternating Heisenberg chain with:
\begin{eqnarray}
H=\sum_{n}(J_1 {\bf S}_{2n}\cdot {\bf S}_{2n+1} + J_2 {\bf S}_{2n+1}\cdot
{\bf S}_{2n+2}).
\end{eqnarray}
with $J_1>0$ and $J_2>0$ (antiferromagnetic couplings) obtained
by the stochastic Quantum Monte Carlo method \cite{Claudius}.
  We define the parameter $\alpha =J_2/J_1$  which measures the ratio
 of the  inter and intra-dimer coupling. $\alpha=0$ corresponds to the dimer
 model.  We introduce also an additional
 parameter $\delta$ so that $J_1=J(1+
\delta)$, $J_2=J(1-\delta)$ and therefore $\alpha=\frac{1-\delta}{1+\delta}$.

\begin{figure}[t]
(a)

\vspace*{5pt}

\centerline{
\epsfig{file=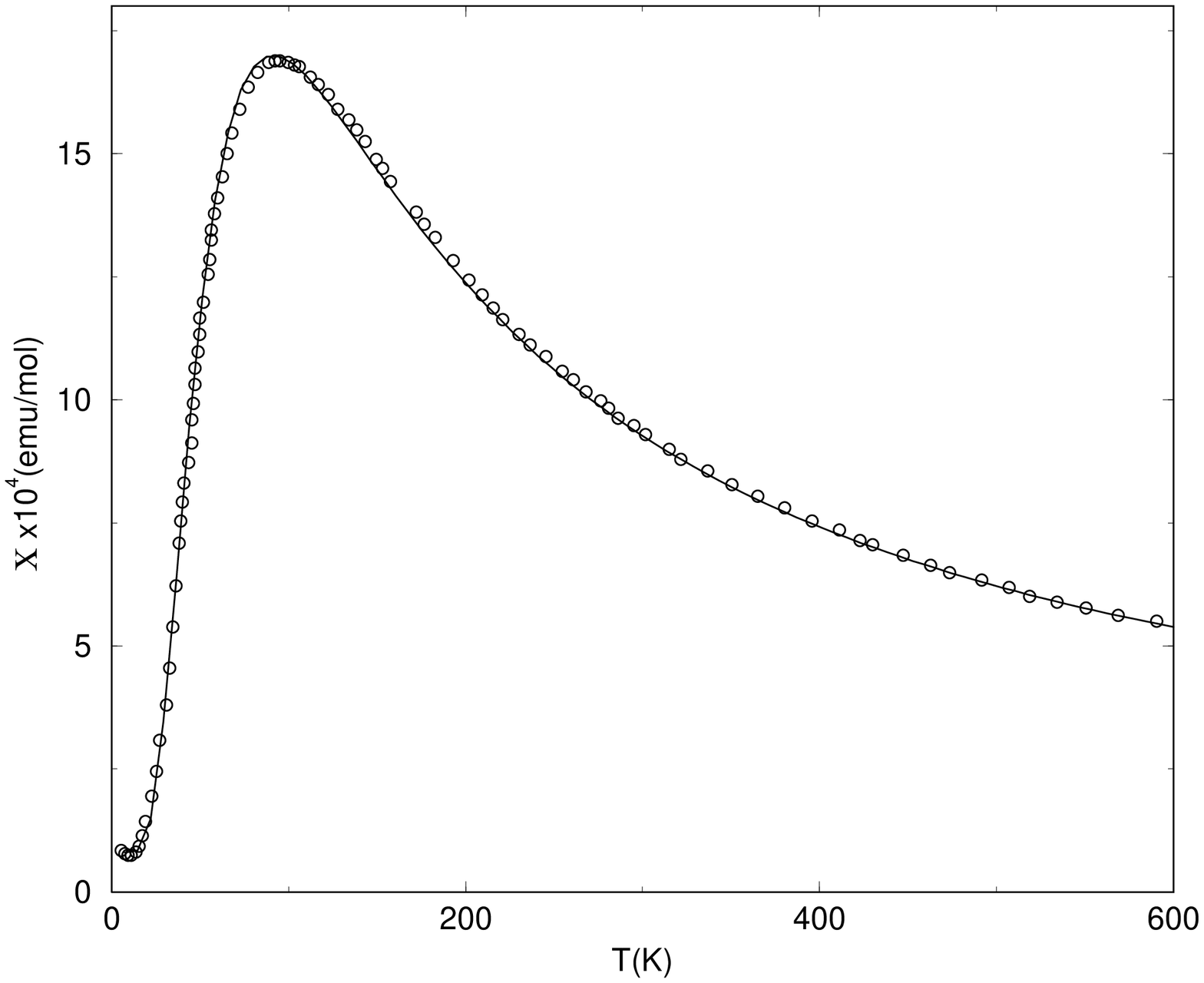,width=0.3\textwidth,angle=-90}
           } 

(b) 

\centerline{
\epsfig{file=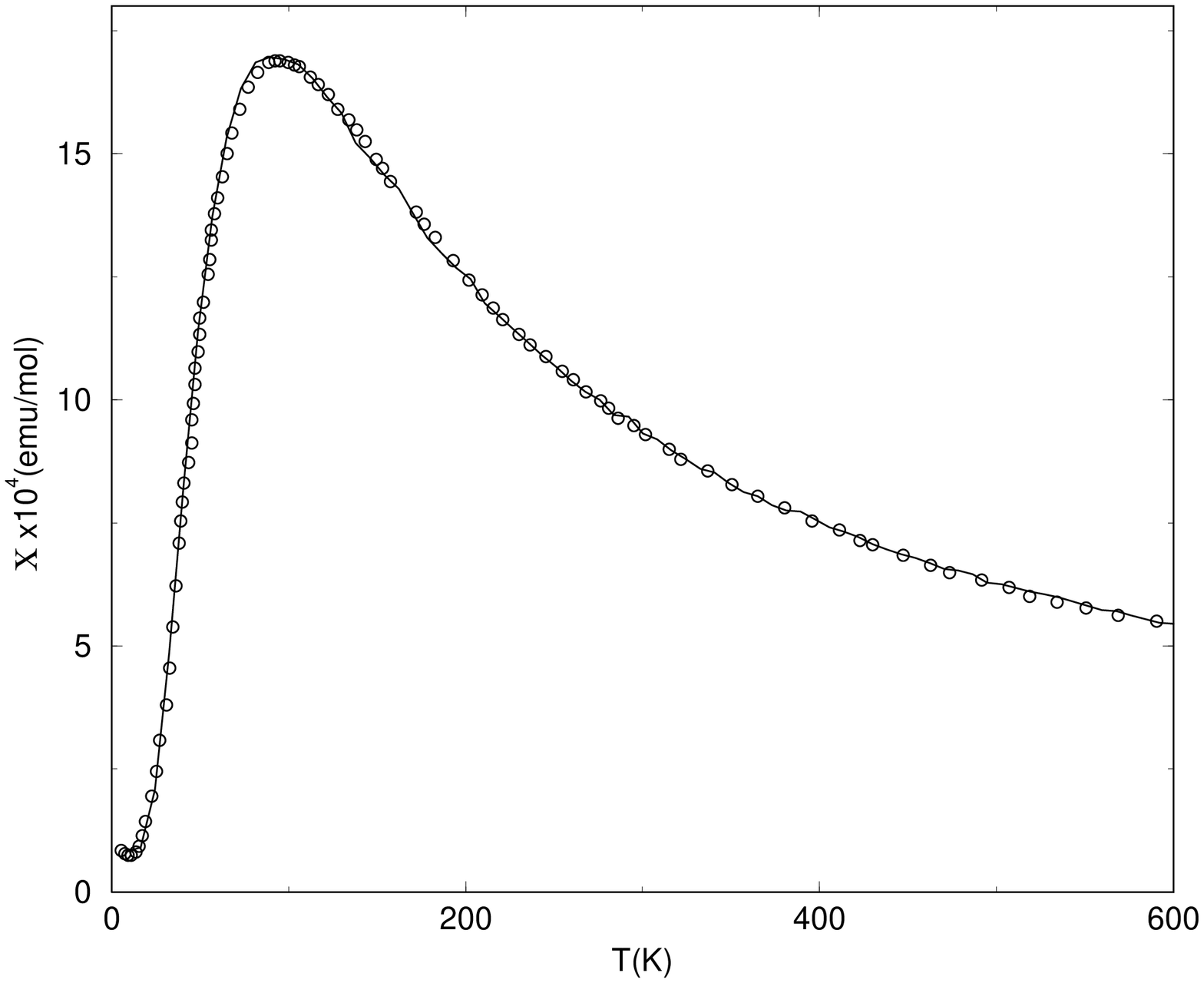,width=0.3\textwidth,angle=-90}
           }

(c)

\centerline{
\epsfig{file=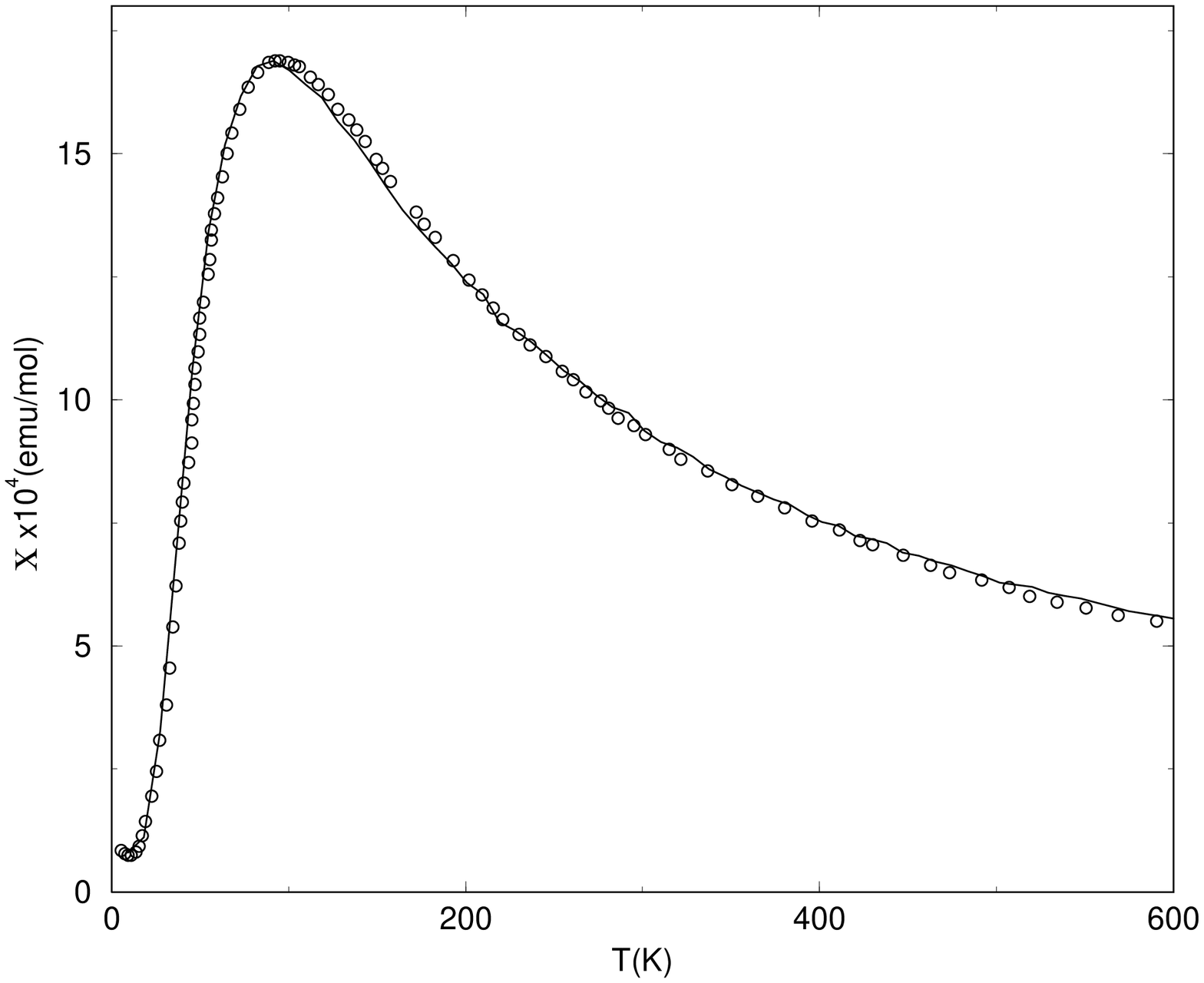,width=0.3\textwidth,angle=-90}
           }
\caption{\label{susceptibility}
Temperature dependence of the magnetic susceptibility of CsV$_2$O$_5$.
 The dots correspond to the experimental data by Isobe {\it et al.}
\protect\cite{Isobe_96_2} where the Curie contribution has been subtracted.
 The solid lines show a fit to (a) spin-$\frac{1}{2}$ Heisenberg dimer
with J/k$_B$=146 K, $g=1.77$, (b)  spin-$\frac{1}{2}$ alternating Heisenberg
chain with $\delta =0.8$, J$_1$/k$_B$=146 K, $g=1.79$ and (c)
 spin-$\frac{1}{2}$ alternating Heisenberg
chain with $\delta =0.6$, J$_1$/k$_B$=146 K, $g=1.81$. }
\end{figure}

In Fig.\ \ref{susceptibility}(a)
 we show the comparison of the experimental data to the
dimer-model susceptibility given by Eq.\ \ref{dimer_susc}.
 We find, in fact, that the best fit
 corresponds to $ J=146K$,  $\chi_0=8\times 10^{-5}{emu}/{mol}$ and
 a value of $g=1.77$, somewhat smaller than the one proposed in
 Ref. \cite{Isobe_96_2}.   Figs.\ \ref{susceptibility} (b) and (c)
  show the comparison
 of the experimental data to the spin-$\frac{1}{2}$
 alternating Heisenberg chain
 model for  $\delta=0.8, \alpha = 0.11$ and $\delta=0.6, \alpha = 0.25$
 respectively.
 We observe that values of 
$\alpha$ up to $0.25$ give still very good fits to the susceptibility
 data by choosing appropriate $g$ and $J$ parameters.  This analysis
 makes plausible our {\it ab initio} results, namely the fact that
 the inter-dimer interactions can be also significant\cite{our}.

 The important
 conclusion that we gain from this susceptibility analysis is that
it is a very hard task to determine the behavior of a compound
only from  susceptibility fits.  Extra information is needed so
 as to minimize the possible options. For instance,  electron spin
 -resonance (ESR) experiments
 could complement the susceptibility measurements by delivering
the value of the gyromagnetic factor. There are in
 the literature other examples ({\it e.g.} VOPO) where the only examination
 of the susceptibility led to a wrong interpretation of the behavior
 of the system\cite{Johnston_87}.

\section{Conclusions}

  We have presented {\it ab initio} band-structure calculations for 
  CsV$_2$O$_5$, a low-dimensional spin-gapped system and member
  of the alkali-vanadate family AV$_2$O$_5$ (A=Na, Li, Cs) which has
 a similar monoclinic structure as the most studied VOPO. 
By means of a {\it downfolding}-tight-binding analysis, we observe
 that the behavior of this system can be best described by a
 spin-$\frac{1}{2}$ alternating chain model with weak interchain interactions.
 This proposal has been complemented by an analysis of 
 available susceptibility  data for this compound. A
 re-examination of the temperature dependence of the experimental
 susceptibility leads us to the conclusion that an alternating chain
 model can also provide a good fitting by choosing appropriate
$g$ and $J$ values.
  We have discussed the structural, electronic and magnetic similarities
and differences between
 CsV$_2$O$_5$ and $\alpha '$-NaV$_2$O$_5$, $\gamma$-LiV$_2$O$_5$ and VOPO.

  With this work we would like to draw the attention on this compound
which shows interesting features, especially since it has a lot
 of points of comparison to the most studied VOPO.   
  We think that new experimental studies like ESR, INS, Raman scattering
 and the study of a possible field-induced magnetic ordering
 under a very strong
 magnetic field
 would be helpful to unveil the
 behavior of this material.
 


\newpage

\vspace*{-0.3cm}

\newpage

\newpage


\begin{thebibliography}{99}
\bibitem{our} A fit of susceptibility data to a model with inter-chain
 couplings hasn't been considered here.
\end{thebibliography}
\end{document}